\documentclass[sigconf]{acmart}

\AtBeginDocument{%
  }

\setcopyright{acmlicensed}
\copyrightyear{2026}
\acmYear{2026}
\acmDOI{XXXXXXX.XXXXXXX}
\acmConference[NLBSE '26]{The 5th International Workshop on
Natural Language-based Software Engineering}{April 12--13,
  2026}{Rio de Janeiro, Brazil }

\acmISBN{978-1-4503-XXXX-X/2018/06}

\newcommand{\etal}{\textit{et al.}}
\usepackage{multirow}

\begin{document}

\title{High-quality data augmentation for code comment classification}

\author{Thomas Borsani}
\orcid{0009-0007-9541-890X}
\affiliation{%
  \institution{Free University of Bozen-Bolzano}
  \city{Bozen-Bolzano}
  \country{Italy}
}
\email{tborsani@unibz.it}

\author{Andrea Rosani}
\orcid{0009-0008-2622-6776}
\affiliation{%
  \institution{Free University of Bozen-Bolzano}
  \city{Bozen-Bolzano}
  \country{Italy}
}
\email{andrea.rosani@unibz.it}

\author{Giuseppe Di Fatta}
\orcid{0000-0003-3096-2844}
\affiliation{%
  \institution{Free University of Bozen-Bolzano}
  \city{Bozen-Bolzano}
  \country{Italy}
}
\email{giuseppe.difatta@unibz.it}

\renewcommand{\shortauthors}{Borsani et al.}

\begin{abstract}
Code comments serve a crucial role in software development for documenting functionality, clarifying design choices, and assisting with issue tracking. 
They capture developers' insights about the surrounding source code, serving as an essential resource for both human comprehension and automated analysis. 
Nevertheless, since comments are in natural language, they present challenges for machine-based code understanding. 
To address this, recent studies have applied natural language processing (NLP) and deep learning techniques to classify comments according to developers’ intentions. 
However, existing datasets for this task suffer from size limitations and class imbalance, as they rely on manual annotations and may not accurately represent the distribution of comments in real-world codebases.
To overcome this issue, we introduce new synthetic oversampling and augmentation techniques based on high-quality data generation to enhance the \textit{NLBSE’26 challenge} datasets. 
Our Synthetic Quality Oversampling Technique and Augmentation Technique (\textbf{Q-SYNTH}) yield promising results, improving the base classifier by $2.56\%$. 
The source code is publicly available at
\href{https://github.com/ThomBors/NLBSE2026}{https://github.com/ThomBors/NLBSE2026}.

\end{abstract}



\begin{teaserfigure}
  \includegraphics[width=\textwidth]{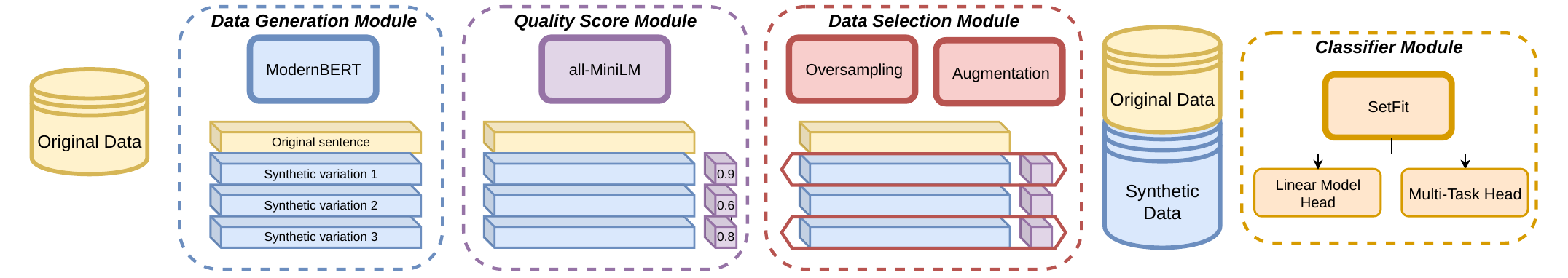}
  \Description{A brief description of the Token Length vs SYNQ Distribution image.}
  \caption{System overview}
  \label{fig:teaser}
\end{teaserfigure}


\maketitle

\section{Introduction}
\label{sec:introduction}
Developers produce various artifacts, including comments, commit messages, and issue tracker entries, which capture their activities and intentions during coding~\cite{ZampettiEtAl2021ComparisonMeans}. 
Among these, code comments are predominantly utilised for their efficiency in documenting and clarifying code choices \cite{ZampettiEtAl2021ComparisonMeans}.
Comments serve multiple roles, with recent research \cite{PascarellaBacchelli2017} aimed at classifying them to understand code elements such as technical debt and possible vulnerabilities~\cite{RussoEtAl2022WeakSATD}.

Developing an automated classifier could support the identification of such problems, reducing manual effort.
To support research in this area, the \textit{NLBSE’26 challenge} \cite{nlbse2026} provides a manually labelled dataset of comments across three programming languages, along with a baseline classification model, STACC \cite{AlKaswan2023}.
The dataset is hindered by two major challenges attributable to the substantial manual effort required for comment labelling. 
First, the sample size is restricted. 
Second, there is an imbalance in class representation that leads to class scarcity. 
These challenges hinder machine learning models, which risk bias towards more prevalent classes or overfitting due to the limited observations available \cite{Japkowicz2002}.

To mitigate these limitations, a commonly used approach is the application of oversampling techniques \cite{Ling1998}. 
This strategy improves the sample size of minority classes by duplicating cases or generating synthetic data \cite{chawla2002smote}. 
However, simple replication does not introduce new information and may induce overfitting as models capture redundant patterns rather than generalisable features \cite{Johnson2019}. 
To address this issue, synthetic approaches such as SMOTE \cite{chawla2002smote}, and its derivate, interpolate between existing minority samples to generate new data, introducing variability while preserving the data structure. 
This method proves effective with structured datasets, where feature interrelations are clear and interpolation maintains semantic coherence. 
Nonetheless, its effectiveness with unstructured or high-dimensional data remains limited \cite{Johnson2019}.

In the era of Generative AI, large language models (\textit{LLMs}) are increasingly being used to generate synthetic data \cite{nakada2025}. 
These models can be used for synthetic oversampling, augmenting minority classes to mitigate class imbalance \cite{Cloutier2023}, as well as for general data augmentation \cite{wang2024}, expanding datasets with diverse and contextually relevant samples. 
Yet an analysis of the quality and representativeness of the synthetic data is not always carried out \cite{Cloutier2023}.

In code comment classification, developers frequently produce remarks that are brief, domain-specific, and grammatically inconsistent.
Accurately reproducing this informal linguistic style presents a substantial challenge for synthetic data generation. 
On one hand, generative models often yield comments that are syntactically correct and semantically refined, thereby deviating from the authentic distribution of real-world data. 
On the other hand, fine-tuning \textit{LLMs} on narrowly defined objectives can undermine reliable deployment, as it can induce unwanted misalignments and behaviour drift \cite{betley2025emergent}.

To tackle these challenges, we present Synthetic Quality Techniques (\textbf{Q-SYNTH}), a synthetic minority oversampling and data augmentation approach that generates diverse variations of the original data while explicitly assessing the diversity and quality of the generated samples. 
We evaluate this method within the \textit{NLBSE’26 challenge} \cite{nlbse2026}.

\section{Synthetic Quality Technique}
The proposed method is composed of three main modules, as illustrated in Figure~\ref{fig:teaser}. 
In this section, we describe the architecture and functionality of each module in detail.

\paragraph{Data Generation Module}
This module is responsible for generating new observations.
To preserve high fidelity with the original data distribution, a BERT-like architecture \cite{devlin-etal-2019-bert} is fine-tuned on the original dataset $(\mathcal{D}_\text{orig})$.
A proportion $\mu$ of tokens in each sequence is randomly masked and the decoder head, trained for masked language modelling, is used to predict plausible token variations.
From the top-$k$ predictions for each masked position, a random subset of $n$ candidates is sampled to ensure stochastic diversity.
Generated sequences $(\mathcal{D}_{\text{gen}})$ are constrained to maintain uniqueness and diversity with respect to the original corpus by monitoring a sequence-matching score $(\delta)$.
During generation, candidates with $\delta$ relative to their source are discarded until the diversity constraint is satisfied.
This procedure ensures that $\lvert \mathcal{D}_{\text{gen}} \rvert = n \times \lvert \mathcal{D}_{\text{orig}} \rvert$, producing a synthetic corpus that expands the original dataset while maintaining the representational variety.

\paragraph{Quality Scoring Module}
This module assigns a quantitative quality score to each synthetically generated observation in $\mathcal{D}_{\text{gen}}$.
Although sequence-matching metrics capture lexical similarity, they fall short in assessing semantic divergence from $\mathcal{D}_{\text{orig}}$.
To evaluate semantic consistency, a pre-trained sentence transformer \cite{reimers-2019-sentence-bert}, optimised and fine-tuned for semantic similarity tasks, provides a quantitative score of semantic fidelity.
Each generated sample $\hat{s}_i \in \mathcal{D}_{\text{gen}}$ is evaluated against its corresponding original sentence $s_i \in \mathcal{D}_{\text{orig}}$, yielding a normalised score $q_i \in [0,1]$ that quantifies the quality of $\hat{s}_i$ relative to $s_i$.

\paragraph{Data Selection Module}
Synthetic samples are drawn from $\mathcal{D}_{\text{gen}}$ using two distinct strategies.
The Oversampling strategy enforces a user-defined quality threshold $QSYNT$ while simultaneously promoting class balance. 
Let $\mathcal{C}$ be the set of classes, $q_i$ the quality score of $\hat{s}_i \in \mathcal{D}_{\text{gen}}$, and $N_c$ the desired number of samples for class $c$. 
The resulting high-quality, class-balanced subset is given by equation~\ref{eq:ovesampling} where $y_i$ denotes the class label associated with $\hat{s}_i$.

\begin{equation}
\label{eq:ovesampling}
\mathcal{D}'_{\text{gen}} = \bigcup_{c \in \mathcal{C}} 
\left\{ \hat{s}_i \in \mathcal{D}_{\text{gen}} \;\middle|\; y_i = c \ \wedge\ q_i \geq QSYNT \right\},
\end{equation}

The Augmentation strategy relaxes the cardinality constraint of equation~\ref{eq:ovesampling}, yielding a restricted synthetic augmentation regime in which the number of selected synthetic samples is not fixed, while evaluation is still performed across the same range of quality thresholds.
The final synthetic-oversampled dataset $\mathcal{D}_{\text{SyO}}$ is defined as:
\(\mathcal{D}_{\text{SyO}} = \mathcal{D}_{\text{orig}} \cup \mathcal{D}'_{\text{gen}}.\)
\section{Computational Experiments}
This section provides details on the experiments conducted to compare the performance of the baseline classifier STACC~\cite{AlKaswan2023} with and without the proposed \textbf{Q-SYNTH} method and its variant for data augmentation.

\begin{figure}
    \centering
    \includegraphics[width=\linewidth]{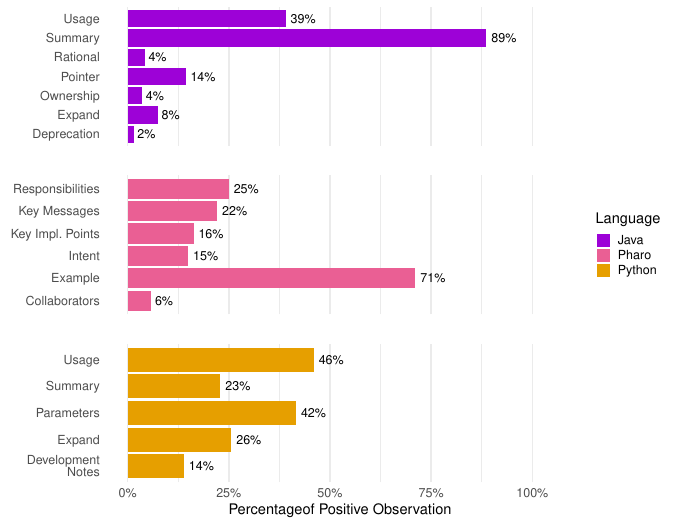}
    \caption{Comparison of percentage of positive observation across categories in the \textit{NLBSE’26 Challenge} dataset \cite{nlbse2026} and variation produced with the Oversampling technique.}
    \label{fig:positiveRatio}
\end{figure}

\begin{figure*}[t!]
    \centering
    \includegraphics[width=\linewidth]{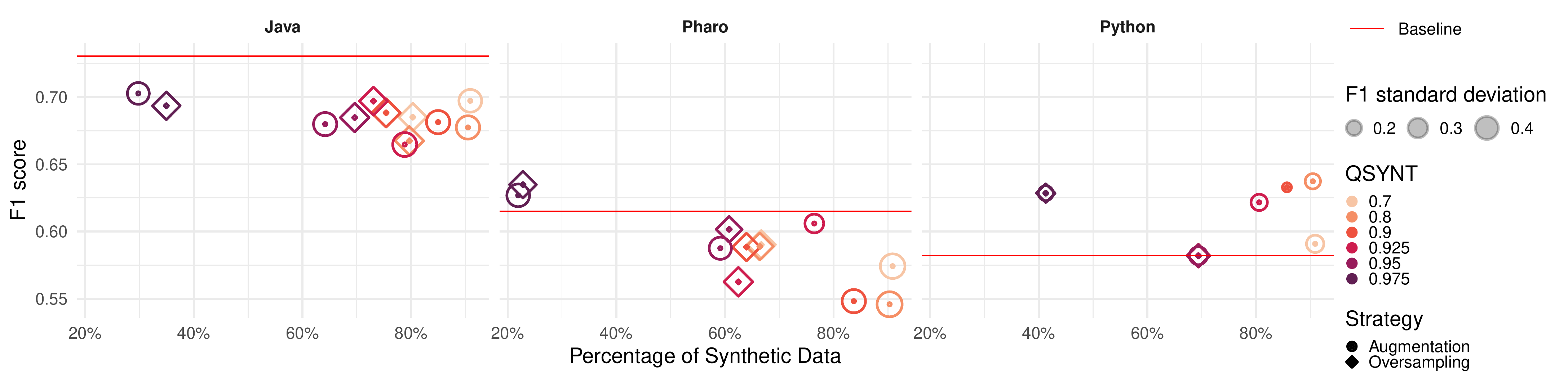}
    \caption{Mean F1 score of the baseline model versus the percentage of synthetic data introduced with two alternative strategies (Augmentation or Oversampling). Point size encodes the relative standard deviation. Colors indicate the selected $QSYNT$ level. The red horizontal lines mark the baseline model performance without synthetic data.}
    \label{fig:f1vsbalancer}
\end{figure*}
\subsection{Dataset}
\label{sec:dataset}
The \textit{NLBSE’26 Challenge} dataset \cite{nlbse2026} is derived from the corpus originally introduced by Rani \etal~\cite{rani2021}. It comprises software comments, file names, and multi-label class vectors for Java, Python, and Pharo. The annotation scheme defines 7 categories for Java comments, 5 for Python, and 6 for Pharo.  
Each comment is assigned to at least one category. The corpus size varies by language, with 6595 instances for Java, 1658 for Python, and 1108 for Pharo. 
Figure~\ref{fig:positiveRatio} reports the proportion of positive labels per category. 
 
In this work, we adopt the \texttt{combo} attributes as the code comment representation and employ the provided 20\% test split solely for the final evaluation of our classifiers.

\subsection{Oversampling Specification}
In the \textbf{Q-SYNTH} methodology, we employ \textit{ModernBERT}~\cite{modernbert} as the sequence transformer within the generation module. 
The model is fine-tuned on the training set with a batch size of 64, a learning rate of $2e-5$, a weight decay of 0.01, and trained for $5$ epochs using a masked language modelling probability of 0.15.

During augmentation, each original instance $s_i$ produces $n = 10$ synthetic variants by randomly sampling from the top-$k = 20$ candidate sentences $\hat{s}_i$, where each candidate is obtained by masking $\mu = 0.25$ of its tokens. 
A diversity threshold of $\delta > 0.95$ is applied to maintain high variability among samples.

The quality scoring component employs the \textit{all-MiniLM-L6-V2} pre-trained model~\cite{hf_allminilm_l6_v2} to compute the quality score $q$ of each synthetic sample. 
In the balancing module, the focus is on augmenting minority classes; however, overgeneration of synthetic samples can cause the model to overfit synthetic patterns rather than real data patterns. 
Thus, to mitigate overfitting risks, the proportion of synthetic data is intentionally constrained to $n = 10$.

\subsection{Experimental Setting}
We assess the impact of the proposed \textbf{Q-SYNTH} oversampling and augmentation techniques on the \textit{NLBSE’26 Challenge} dataset by comparing the performance of the STACC baseline classifier~\cite{AlKaswan2023} with and without their application.
To analyse how the quality criterion affects performance, we evaluate \textbf{Q-SYNTH} under multiple quality thresholds,
$Q_{\text{SYNTH}} \in \{0.70, 0.80, 0.90, 0.925, 0.95, 0.975\}$.


\section{Results}

\begin{figure}[thbp]
    \centering
    \includegraphics[width=0.8\linewidth]{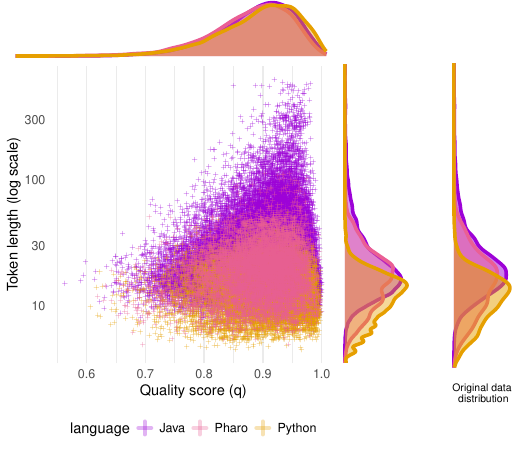}
    \Description{A brief description of the Token Length vs SYNQ Distribution image.}
    \caption{Distribution of generated sentence token lengths produced by the SetFit tokenizer and their corresponding quality scores ($q$). The central plot shows the joint distribution, while the marginal density plots display the individual distributions of token length and quality score.}
    \label{fig:distribution}
\end{figure}

This section presents the experimental findings. 
Using a masking ratio of $\mu = 0.25$ on the original sentences yields sufficient diversity in the resulting text. As shown in Figure~\ref{fig:distribution}, our method produces synthetic data with an average quality score of roughly $0.9$, exhibiting a long-tailed distribution with a smooth decrease toward lower $q$ values and only a few samples below $0.7$.

The token-length distribution of the synthetic sentences closely matches that of the original corpus.  
Nevertheless, longer sentences $s_i$ tend to achieve higher $q$ scores, as also shown in Figure~\ref{fig:distribution}.

The dataset contains a large number of sentences; however, most are approximately 20 tokens long and show limited lexical variability (Figure~\ref{fig:distribution}). 
To preserve data quality and diversity, the number of generated observations must, therefore, be strictly constrained. As a consequence, the pool of high-quality variants is rapidly depleted, since each sentence allows only a narrow scope for meaningful synthetic modification. 
This behavior is shown in Figure~\ref{fig:f1vsbalancer}, which shows that as $QSYNT$ increases, the fraction of synthetic data declines.
Around $QSYNT \approx 0.975$, synthetic samples yield only minor improvements in class balance, whereas reducing this threshold produces more pronounced gains in class balance. 

With high-quality synthetic data, the F1 score exceeds the baseline for both Pharo and Python (Figure~\ref{fig:distribution}). 
Oversampling, while only modestly improving class balance, yields higher F1 scores than augmentation at comparable quality levels. For Java, the performance difference between augmentation and oversampling is smaller, and neither approach improves upon the baseline. In contrast, incorporating lower-quality synthetic data is detrimental:
at $QSYNT \approx 0.7$, more than 80\% of the training data consists of low-quality synthetic samples, introducing substantial noise into the learning process.

\section{Final Model Architecture}
STACC \cite{AlKaswan2023} enhanced 
with Q-SYNTH did not consistently improve upon the baseline performance.
This limitation is likely attributable to the amount of synthetic data, which was insufficient to fully balance the dataset and, consequently, constrained classification performance.

To address this issue, we replace the linear regression head originally adopted in STACC  \cite{AlKaswan2023} with a fully connected multi-task learning (MTL) head optimised using the SAMGS \cite{borsani2025} optimiser.

Previous work \cite{Moritz2025} has shown that dynamic loss-balancing strategies, such as SAMGS, tend to be unstable under conditions of severe data scarcity. In our approach, although synthetic data generation does not fully resolve class imbalance, it sufficiently alleviates data scarcity to enable the effective deployment of a multi-task optimisation framework.

Consequently, the proposed architecture achieves improved overall performance and enhanced stability across all evaluated tasks, as reported in Table~\ref{tab:finalNLBSE2026}. 
\begin{table}[thbp]
    \centering
    \caption{Comparison of results with $\Delta$ F1\textsubscript{c} indicates the difference between our best configuration and the baseline.}
    \label{tab:finalNLBSE2026}
    \resizebox{\linewidth}{!}{ 
        \begin{tabular}{lllllllllll}
            \hline
             &  & \multicolumn{3}{c}{\multirow{2}{*}{Baseline (STACC)}} &  & \multicolumn{3}{c}{\multirow{2}{*}{Best solution}} & &\\
             &  & \multicolumn{3}{c}{} &  & \multicolumn{3}{c}{} & &\\ \hline
             & Class label & P\textsubscript{c} & R\textsubscript{c} & F1\textsubscript{c} &  & P\textsubscript{c} & R\textsubscript{c} &  F1\textsubscript{c} & & $\Delta$ F1\textsubscript{c} \\ \hline
             &  & \multicolumn{3}{c}{} &  \multicolumn{5}{c}{MTL - OVER. - 0.975} \\ \hline
            \multirow{7}{*}{\rotatebox[origin=c]{90}{Java}} & Summary & 87.12 & 88.67 & 87.89 & & 87.24 & 89.64 & 88.43 & &\phantom{-}0.54\\                 
            & Ownership & 100 & 100 & 100 & & 100 & 100 & 100 & &\phantom{-}0\\                              
            & Expand & 33.01 & 43.04 & 37.36 & & 40.3 & 34.18 & 36.99 & &-0.37\\                  
            & Usage & 88.38 & 85.08 & 86.7 & & 86.01 & 83.39 & 84.68 & &-2.02\\
            & Pointer & 77.56 & 96.8 & 86.12 & & 77.16 & 100 & 87.11 & &\phantom{-}0.99\\
            & Deprecation & 87.5 & 70 & 77.78 & & 77.78 & 70 & 73.68 & &-4.10\\
            & Rational & 31.17 & 41.38 & 35.56 & & 55.26 & 36.21 & 43.75 & &\phantom{-}8.19\\\hline
             & Java Average & 72.11 & 75 & 73.06 & & 74.82 & 73.35 & 73.52 & & \phantom{-}0.46\\\hline
             &  & \multicolumn{3}{c}{} & \multicolumn{5}{c}{Linear - AUG. - 0.80} \\ \hline
            \multirow{5}{*}{\rotatebox[origin=c]{90}{Python}} & Usage & 66.67 & 68.13 & 67.39 & & 72.22 & 71.43 & 71.82 && \phantom{-}4.43\\                    
            & Parameters & 68.82 & 75.29 & 71.91 & & 74.16 & 77.65 & 75.86 && \phantom{-}3.95\\               
            & Dev. Notes & 21.92 & 50 & 30.48 & & 38.1 & 50 & 43.24 && \phantom{-}12.76\\             
            & Expand & 45.33 & 66.67 & 53.97 & & 60.71 & 66.67 & 63.55 && \phantom{-}9.58\\                   
            & Summary & 68.97 & 65.57 & 67.23 & & 72.92 & 57.38 & 64.22 && -3.01\\ \hline
             & Python Average & 54.34 & 65.13 & 58.2 & & 63.62 & 64.63 & 63.74 & &\phantom{-}5.54\\ \hline
             &  & \multicolumn{3}{c}{} &  \multicolumn{5}{c}{MTL - AUG. - 0.9} \\ \hline
            \multirow{7}{*}{\rotatebox[origin=c]{90}{Pharo}} & Key Imp. Points & 56.25 & 64.29 & 60 & & 63.33 & 67.86 & 65.52 & &\phantom{-}5.52\\
            & Example & 88.64 & 87.64 & 88.14 & & 85.23 & 84.27 & 84.75 & &-3.39\\
            & Responsibilities & 63.27 & 73.81 & 68.13 & & 64.86 & 57.14 & 60.76 & &-7.37\\
            & Intent & 72 & 85.71 & 78.26 & & 61.54 & 76.19 & 68.09 & &-10.17\\
            & Key Mess. & 47.83 & 73.33 & 57.89 & & 58.06 & 60 & 59.02 & &\phantom{-}1.13\\
            & Collaborators & 10.34 & 42.86 & 16.67 & & 50 & 42.86 & 46.15 & &\phantom{-}29.48\\\hline
             & Pharo Average & 56.39 & 71.27 & 61.52 & & 63.84 & 64.72 & 64.05 & &\phantom{-}2.53\\ \hline
             & \textbf{Grand Average} & 61.93 & 71.02 & 65.08 & & 68.05 & 68.05 & 67.65 & &\phantom{-}2.56\\ \hline
        \end{tabular}
    }
\end{table}

\section{Conclusion}
Synthetic data enhances classification performance, with oversampling yielding the largest gains for Pharo and Python. Across all languages, data quality consistently outweighs quantity: compact, high-quality synthetic datasets outperform larger but noisier alternatives. These findings indicate that quality-aware synthetic data generation is an effective strategy for mitigating data scarcity and class imbalance in code comment classification. Future work will focus on identifying and selecting generated samples with maximal impact on downstream performance. The final score achieved in the NLBSE’26 challenge was 0.73.
\begin{acks}
This publication has received funding from the European Union’s HORIZON Research and Innovation Actions through the project AI4SWEng (Agreement No 101189908).
\end{acks}

\bibliographystyle{ACM-Reference-Format}
\bibliography{ref}

\end{document}